\documentclass[conference]{IEEEtran}
\IEEEoverridecommandlockouts
\usepackage{cite}
\usepackage{amsmath,amssymb,amsfonts}
\usepackage{algorithmic}
\usepackage{graphicx}
\usepackage{textcomp}
\usepackage{xcolor}
\def\BibTeX{{\rm B\kern-.05em{\sc i\kern-.025em b}\kern-.08em
    T\kern-.1667em\lower.7ex\hbox{E}\kern-.125emX}}
\begin{document}

\title{Learning and Unlearning: \\Bridging classification, memory and generative modeling in Recurrent Neural Networks\\

\author{\IEEEauthorblockN{Enrico Ventura}\\
\IEEEauthorblockA{\textit{Bocconi Institute for Data Science and Analytics (BIDSA)}} 
\textit{Bocconi University}\\
Milan, Italy.\\
enrico.ventura@unibocconi.it}

}

\maketitle

\begin{abstract}
The human brain is a complex system that is fascinating scientists since a long time. Its remarkable capabilities include categorization of concepts, retrieval of memories and creative generation of new examples. At the same time, modern artificial neural networks are trained on large amounts of data to accomplish these same tasks with a considerable degree of precision.
By contrast with biological systems, machines appear to be either significantly slow and energetically expensive to train, suggesting the need for a paradigmatic change in the way they learn. 
We here review a general learning prescription that allows to perform classification, memorization and generation of new examples in bio-inspired artificial neural networks. The training procedure can be split into a prior Hebbian learning phase and a subsequent anti-Hebbian one (usually referred to as Unlearning). The separation of training in two epochs allows the algorithm to go fully unsupervised while partially aligning with some modern biological theories of learning. 
\end{abstract}


\section{Introduction}
One of the major challenges for the scientific community is to create artificial machines that are closer to biological information processing systems, both in the way they learn and in the topological structure. Achieving this goal would introduce more efficient technological devices as well as boosting a deeper comprehension of the brain
functioning. Statistical mechanics, a branch of physics that deals with complex systems, has made significant
contributions to the study of both artificial and biological neural networks by modeling them as simple graphs of
interconnected units with adjustable interactions \cite{amit_modeling_nodate}. Several algorithms have been proposed to optimize the neural
connections to enable the network to mimic real brain functions. Among the first contributions to this field there
is the Hebbian rule. This learning model gives a valuable prescription for the synaptic connections to store a number of stimuli up to a limited degree of corruption \cite{hebb_organization_1949,hopfield_neural_1982}. Further studies, such as
Gardner’s \cite{gardner_space_1988}, introduced the idea of changing synaptic connections to optimize memory storage,
while Hinton and coworkers showed that a similar neural network can effectively learn distributions of large
amounts of external stimuli and create new ones from the neural dynamics \cite{hinton_unsupervised_1999}. 
The goal of this paper is to delineate a particular learning prescription that can perform different tasks through a slight change in the control parameters and parameter initialization. Specifically, we will analyze the following tasks: classifying unseen data-points (i.e. \textit{classification}); retrieving specific stimuli from corrupted examples (i.e. \textit{associative memory}); generating new examples being consistent with past received stimuli (i.e. \textit{generative} task).
The general idea behind this method, which is already present in the literature under different names and forms, is matching correlations in the neural activity with ones measured from stimuli/data-points (i.e. it is a \textit{moment-matching} algorithm). This concept is at the very basis of energy-based generative models, and corroborates the idea that classification and memorization may be only a particular case of a generation.\\
This idea is carried out along the paper in the following order. 
Section \ref{sec:RNN} describes the architecture of the neural network that we are going to use, as well as the network dynamics.
In Section~\ref{sec:class} we introduce a first version of our moment-matching algorithm showing a recurrent neural network that classifies examples. Original results about this approach will be contained in \cite{prep}. In Section~\ref{sec:memory} some assumptions over the previous prescription are
relaxed towards an algorithm that is more biologically relatable. The memorization algorithm was introduced by Hopfield in \cite{hopfield_unlearning_1983} and it consists in a Hebbian
initialization of the synaptic connections followed by a anti-Hebbian Learning of an ensemble of stimuli. 
This procedure is generally known in the literature under the name of Unlearning.
It will be evident that a network trained
according to this method can memorize examples in a more robust fashion with respect to the Hebbian rule alone,
reaching a performance that is comparable with the best associative memory models in literature \cite{gardner_phase_1989,gardner_optimal_1989}. The results of our analysis are gathered in publications \cite{benedetti_supervised_2022, bene2}. 
Successively, in Section~\ref{sec:gene} we will test the algorithm on a larger amount of data in order to show that it can very well learn a
model that approximates their probability distribution emulating the
Boltzmann Machine mechanism \cite{hinton_unsupervised_1999}. More insights about the use of the Unlearning rule for the generative task are reported in \cite{ventura_2024}. 
The Conclusion of the paper will present some open problems left by our investigation, highlighting possible useful contributions to the domains of Biology, Computer Science and Physics.

\section{\label{sec:RNN} Recurrent Neural Networks}
We will thus model our neural network as a fully connected network of $N$ binary neurons $\{S_i = \pm 1\}_{i=1}^N$, linked by couplings $J_{ij}$. This model resembles enters the class of disordered binary networks studied by statistical mechanics \cite{peretto_collective_1984, mezard_spin_1986}. In this work we will consider the matrix $\boldsymbol{J}$ to be real symmetric with null diagonal elements. Generally speaking, the network configuration $\vec{S}$ evolves according to the following stochastic dynamics
\begin{equation}
    \label{eq:stoch_dynamics}
    P_{T}^{(t+1)}(S_i) = \frac{e^{\frac{1}{T}S_i h_i^{(t)}}}{e^{\frac{1}{T}S_i h_i^{(t)}}+e^{-\frac{1}{T}S_i h_i^{(t)}}},
\end{equation}
where the configuration of each neuron $S_i^{(t)}$ is sampled sequentially in $t$, in a random order, according to the probability $P_T^{(t)}(S_i)$, $T$ is a temperature parameter controlling the degree of stochasticity, and $h_i^{(t)}$ is the local field to neuron $S_i^{(t)}$, defined as
\begin{equation}
    h_i^{(t)} = \sum_{j\neq i}J_{ij}S_j^{(t)}.
\end{equation}
When $T \rightarrow 0$ the dynamics reduces to the following deterministic rule
\begin{equation}
    \label{eq:dynamics}
    S_i^{(t+1)} = \text{sign}\left(h_i^{(t)}\right).
\end{equation}
When $\boldsymbol{J}$ is \textit{symmetric} the model reaches an equilibrium state for any $T$, which is a global minimum of the \textit{free-energy} function of the system
\begin{equation}
\label{eq:free_ene}
    F_{T}[\boldsymbol{J}] = -T\log{\left( \sum_{\vec{S}}e^{-\frac{1}{T} E[\vec{S}|\boldsymbol{J}]}\right)},
\end{equation}
where
\begin{equation}
    E[\vec{S}|\boldsymbol{J}] = -\frac{1}{2}\sum_{i,j}S_i J_{ij}S_j,
\end{equation}
is called $\textit{energy}$.
In the deterministic scenario described by Eq.~\ref{eq:dynamics}, the steady state will be a single stable fixed point of the dynamics, $\vec{S}^*$ also called \textit{attractor}, namely a local minimum of the \textit{energy} function.\\ \\
In both biological and artificial neural networks, synaptic connections $\boldsymbol{J}$ can be trained by the effect of external stimuli from the environment to learn to perform a task. 
We now propose a quite general training prescription so that the model can learn how to \textit{classify} stimuli, \textit{memorize} them, and creatively \textit{generate} new ones, by only means of the dynamics expressed in Eq.~\ref{eq:stoch_dynamics}.
\section{\label{sec:class}Classification}
In all classification problems there is a set of $N$ dimensional data-points $\{\vec{S}^{a}\}_{a=1}^M$ that have to be assigned to $p$ possible classes. A \textit{classifier} is a model that has learned a function $\phi_J$ such that
\begin{equation}
    \phi_J\left(\vec{S}^a \right)=\sigma^a,\hspace{0.5cm}\sigma^{a} \in \{\text{class 1},\text{class 2},...,\text{class p}\}.
\end{equation}
with $\boldsymbol{J}$ being the trainable parameters of the model.
While some classifiers are trained on a training-set $\{\vec{S}^a,\sigma^a\}_{a = 1}^M$ with previously assigned labels \cite{gardner_space_1988}, other models infer the classes directly from the data \cite{bengio}.\\
Let us consider the case where the classes are encoded in a $N$ dimensional collections of features, i.e. $\vec{\sigma}^{a}\in \{-1,+1\}^N$ and we are provided with $p$ vectors $\{\vec{\xi^{\mu}}\}_{\mu = 1}^p$ that play the role of prototypes for the classes. 
Then the problem translates into finding a function $\vec{\phi}_J$ such that
\begin{equation}
    \label{eq:percept_class_vec}
    \small
    \vec{\phi}_J\left(\vec{S}^{a} \right)=\vec{\sigma}^{a} ,\hspace{0.5cm}\vec{\sigma}^{a}\in\{\vec{\xi}^1,\vec{\xi}^2,..,\vec{\xi}^p\},\hspace{0.5cm}a=1,..,M.
\end{equation} 
We will now show that this problem can be solved through a recurrent neural network with a coupling matrix $\boldsymbol{J}$ with the network dynamics playing the role of the classifying function $\vec{\phi}_J$.

\subsection*{Recurrent Neural Network-based solution}
From the hypothesis of the classification problem, we are provided with a set of $p$ vectors $\{\vec{\xi^{\mu}}\}_{\mu = 1}^p$ where $\vec{\xi^{\mu}}\in\{-1,+1\}^N$ that represent the prototypes of the \textit{classes}. The number of classes is chosen to be $p = \alpha N$, i.e. extensive in the dimension of the data-set.
Let us now consider the neural network model described in Section~\ref{sec:RNN}. The final goal of our procedure will consist into training a neural system that has the classes as the only $p$ attractors of the dynamics, having connected basins of attraction spanning the entire space of configurations. 
We now define two matrices that will enter the training of the parameters. The covariance matrix evaluated across the patterns is a real symmetric matrix $\boldsymbol{W}\in \mathbb{R}^{N\times N}$ defined as
\begin{equation}
\label{eq:W}
    \boldsymbol{W} = \frac{1}{p}\sum_{\mu=1}^p\vec{\xi^{\mu}}(\vec{\xi^{\mu}})^{\intercal} .
\end{equation}
We will generally refer to this matrix as \textit{Hebbian} matrix. 
Let us consider the system to evolve through the $T \rightarrow 0$ dynamics in Eq.~\ref{eq:dynamics}, with vectors $\vec{S}^*$ being the stable fixed points, i.e. the local minima of the energy function. 
Hence, we call $\boldsymbol{C}\in \mathbb{R}^{N\times N}$ the covariance matrix over all the local minima $\vec{S}^*$, i.e. 
\begin{equation}
    \label{eq:fp_ave}
    \boldsymbol{C} = \frac{1}{\Omega}\sum_{*}^{\Omega}\vec{S}^* (\vec{S}^*)^{\intercal},
\end{equation}
where $\Omega$ is the total number of local minima of the energy associated to a realization of $\boldsymbol{J}$. 
\\We now present a simple recursive procedure that trains a recurrent neural network that classifies: 
\begin{itemize}
    \item Initialize $\boldsymbol{J}^{(d=0)}$ in an arbitrary fashion that preserves the symmetry. Then, for each iteration $d>0$: 
    \begin{enumerate}
        \item Compute $\boldsymbol{C}^{(d)}$. Stop the algorithm if $\boldsymbol{C}^{(d)}=\boldsymbol{W}$. Otherwise:
        \item Update $\boldsymbol{J}^{(d)}$ as:
        \begin{equation}
            \label{eq:up_J}
            \boldsymbol{J}^{(d+1)} = \boldsymbol{J}^{(d)}+\frac{\lambda}{N}\left(\boldsymbol{W}-\boldsymbol{C}^{(d)} \right),
        \end{equation}
        with $\lambda >0$ and $J_{ii} = 0$, $\forall i$.
    \end{enumerate}
\end{itemize} 
Whether the algorithm converges it does in one configuration of $\boldsymbol{J}$ such that the only stable fixed points of Eq.~\ref{eq:dynamics} will be $\{\vec{\xi}^{\mu}\}_{\mu}^p\cup\{-\vec{\xi}^{\mu}\}_{\mu}^p$. 
As a consequence, any initial configuration will be contained in the \textit{basin of attraction} of one of the $2p$ vectors (including their inverse copy). 
The iterative neural dynamics now plays the role of the classifying function $\vec{\phi}_J$, that takes any neural configuration $\vec{S}$ unseen by the model, and gives back the closest class vector. The number of iteration of the dynamics scales as $\mathcal{O}(N^2)$.
\\Numerics suggest that basins of attraction appear to be connected in the neural configuration space, i.e. they gather data-points that are similar with each other. 
On the other hand, the size and the shape of the basins of attraction, as well as the maximum density of classes $\alpha_c$, will depend on the structure of the class vectors themselves \cite{prep}. 
\subsection*{Limitations of the algorithm }
The algorithm here explained conserves its efficacy as long as the matrix $\boldsymbol{C}$ can be computed exactly. This implies the evaluation of all $2^N$ neural configurations, which becomes computationally intractable when the size of the system becomes larger than $N\simeq 25$ on modern machines.
When $N$ is large one has to sample a number of configurations that is polynomial in $N$ to approximate the $\boldsymbol{C}$ matrix. Even though the patterns are still stable fixed points of the dynamics, the injected sampling noise implies a proliferation of an exponential number of spurious attractors, with a shrinking of the basins of attraction. As a consequence, data-points that conserve a significant overlap with the class vectors will still be classified correctly, while points falling out of the basin of attraction will typically belong to the basin of attraction of a spurious state. Even though this aspect has not been studied in detail, we expect some of these states to be only weakly correlated with the classes, leading to possible mis-classification of the data-points. This behaviour is typical of the associative memory models that will be the object of the following Section.
\section{\label{sec:memory}Associative Memory}
With the term \textit{associative memory} we define the capability of the neural system of reconstructing a given concept (i.e. a \textit{memory}) when a corrupted version of it is displayed as an input of the neural dynamics \cite{amit_modeling_nodate}. 
This type of task is formally similar to classification, yet it does not require a strict separation of the neural configuration space into classes and relies on a partial knowledge of the patterns. The memorization problem thus appears as a more biologically relatable counterpart of classification.  
The usual models employed to reproduce memorization are the recurrent neural networks described in Section~\ref{sec:RNN} evolving through the deterministic dynamics in Eq.~\ref{eq:dynamics}. To define the problem we generate a set of $p$ vectors $\{\vec{\xi^{\mu}}\}_{\mu = 1}^p$ where $\vec{\xi^{\mu}}\in\{-1,+1\}^N$ that we call patterns, and play the roles of the memories to be stored. The goal of an associative memory model is to optimize its connections $\boldsymbol{J}$ to stock the patterns into finite sized basins of attraction. At this point, the dynamics in Eq.~\ref{eq:dynamics} can be initialized on a state $\vec{S}^{(0)}$ that has an overlap $m_0$ with one reference pattern $\vec{\xi^1}$, i.e.
\begin{equation}
   m_0 = \lim_{N\rightarrow \infty}\frac{1}{N}\sum_{i = 1}^N \xi_i^1 S_i^{(0)}. 
\end{equation}
The dynamics will subsequently converge on a stable fixed point $\vec{S}^*$ that has a final overlap with the same reference pattern defined as  
\begin{equation}
\label{eq:m}
m_f(m_0) = \lim_{N \rightarrow \infty}\frac{1}{N}\sum_{i = 1}^N \xi_i^1 S_i^*.
\end{equation}
The minimum value of $m_0$ such that $m_f(m_0) = \mathcal{O}(1)$ defines the size the basin of attraction that contains the pattern. 
As long as such basin of attraction exists we are in a \textit{retrieval} phase of the model. \\
On the other hand, we define $\textit{perfect-retrieval}$ as the capability to perfectly retrieve each pattern when the dynamics is initialized on the pattern itself. It is here convenient to define a quantity, called \textit{stability}, defined as
\begin{equation}
    \label{eq:stab}
    \small
    \Delta_i^{\mu} = \frac{\xi_i^{\mu}}{\sqrt{N}\sigma_i}\sum_{j = 1}^N J_{ij}\xi_j^{\mu}, \qquad \sigma_i = \sqrt{\sum_{j=1}^N J_{ij}^2/N}.
\end{equation}
The condition for having perfect-retrieval reads  
\begin{equation}
    \label{eq:perf_ret}
    \Delta_i^{\mu} > k > 0 \hspace{0.4cm}\forall i,\mu,  
\end{equation}
with $k$ called \textit{margin}. \\

\subsection*{Hebbian Learning and Hebbian Unlearning}

\begin{figure}[htbp]
\centering
\includegraphics[width=\linewidth]{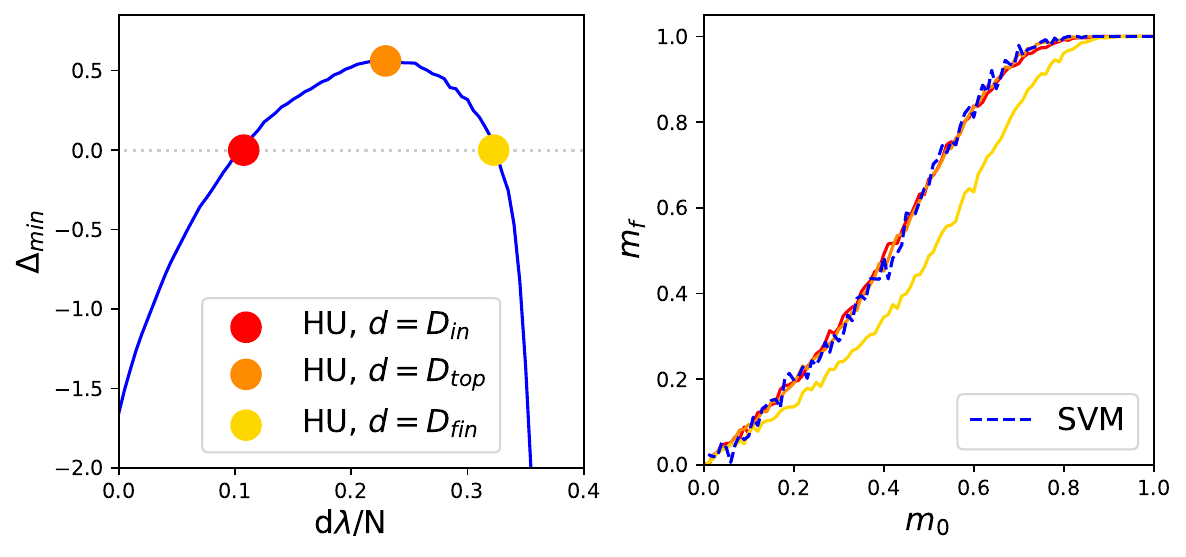}
\caption{Left: The minimum stability $\Delta_{\text{min}} = \text{min}_{i,\mu}(\Delta_i^{\mu})$ as a function of the normalized algorithm time. The threshold $\Delta = 0$ is indicated with the \textit{gray} dotted line. Three relevant amount of iterations are indicated by the colored circles: $d = D_{in}$ in \textit{red}, $d = D_{top}$ in \textit{orange}, $d = D_{fin}$ in \textit{yellow}. All measures are averaged over 50 realizations of the network. Right: Retrieval map $m_f(m_0)$ for a SVM and the unlearning algorithm at the three relevant steps $\{D_{in}, D_{top}, D_{fin}\}$, with the same color code used on the plot on the left panel. All measures are averaged over 10 realizations of the network. Choice of the parameters: $N = 100$, $\alpha = 0.3$, $\lambda = 10^{-2}$. Image from \cite{bene2}.}
\label{fig:unl}
\end{figure}

We call \textit{Hebbian learning} the choice of $\boldsymbol{J} = \boldsymbol{W}$, with $\boldsymbol{W}$ defined according to Eq.~\ref{eq:W}. This building prescription for the $\boldsymbol{J}$ allows to memorize and retrieve up to $p \simeq 0.14 N$ patterns.  
Up to this threshold the system lives in a retrieval phase and $m_f(1) \geq 0.97$ \cite{amit_storing_1985}. \\
It has been showed that one can substantially improve the retrieval performance of the network by combining Hebbian learning with a second learning rule named \textit{Hebbian unlearning} (or just \textit{unlearning}) \cite{hopfield_unlearning_1983}. Specifically, we can enter a phase of perfect retrieval up to higher capacities while notably increasing the size of the basins of attraction. The full learning prescription for the couplings reads:
\begin{itemize}
    \item Initialize by Hebbian Learning, i.e. $$\boldsymbol{J}^{(0)}=\boldsymbol{W}.$$ 
    \item Reiterate the following routine:
    \begin{enumerate}
        \item Run the dynamics~\ref{eq:dynamics} to sample $\Omega$ stable fixed points $\vec{S}^{*}$ and compute
        \begin{equation}
            \Tilde{C}_{ij}^{(d)} = \frac{1}{M}\sum_{*}^{\Omega}S_i^* S_j^*.
        \end{equation}
    \item Update couplings as
    \begin{equation}
        \label{eq:unl_rule}
        \small
        \boldsymbol{J}^{(d+1)} = \boldsymbol{J}^{(d)} -\frac{\lambda}{N}\boldsymbol{\Tilde{C}}^{(d)},
    \end{equation}
    with $\lambda >0$ and $J_{ii} = 0$, $\forall i$.
    \end{enumerate}
\end{itemize}
From now on we will consider the $\Omega = 1$ case, since it is the most widely studied in the literature. 
However, considering $\Omega\gg 1$ mainly affects the learning time scale, without any proper consequence on the thermodynamics of the model. 
As another difference from our presentation of the algorithm, previous works initialize the coupling matrix as $\boldsymbol{J}^{(0)}=\alpha\boldsymbol{W}$, but this has the only consequence of rescaling the learning rate as $\lambda/N \rightarrow \lambda/p$.\\
Figure~\ref{fig:unl} shows the minimum stability as a function of the learning time steps, for one choice of the control parameters $\alpha, N, \lambda$. One can see that perfect retrieval is reached at $d = D_{in}$ and conserved until $d = D_{fin}$. The perfect-retrieval phase of the model is present up to $\alpha_c \simeq 0.6$. 
\\We can compare the performance of this procedure with the recurrent version of a linear perceptron that maximizes its margin, i.e. the recurrent neural network whose couplings $\boldsymbol{J}$ satisfy the following conditions over the patterns
\begin{equation}
    \Delta_i^{\mu} > k_{max}(\alpha) \hspace{0.4cm}\forall i,\mu,  
\end{equation}
where $k_{max}$ is the highest value of the margin that one can afford at a given $\alpha$ \cite{gardner_space_1988,gardner_optimal_1989,brunel_is_2016}. The solution $\boldsymbol{J}_{\text{SVM}}$ that satisfies such conditions is unique for any $\alpha$, and we will call the resulting network \textit{(Recurrent) Support Vector Machine}, or just SVM \cite{battista_capacity-resolution_2020}. 
This network has, by construction, very robust patterns, that are perfectly retrieved and have very large basins of attraction. The critical capacity for this model is the same as the linear perceptron, i.e. $\alpha_c = 2$. 
As showed by the right panel in Figure~\ref{fig:unl} the basins of attraction obtained through the Unlearning procedure after $D_{in}$ learning steps are equivalent in size to the ones reached a SVM. 
This property holds in the large $N$ limit, at least for $\alpha < 0.6$ as displayed in \cite{benedetti_supervised_2022,van_hemmen_increasing_1990}. 
By contrast with SVMs, the Unlearning algorithm is fully unsupervised: the memories are seen only once, at initialization, while the rest of training only relies on the structure of the stable fixed points. 
\subsection*{Unlearning as a noisy Perceptron}
The Unlearning rule can be derived from a linear perceptron that learns from noisy versions of the data-points. This noise-based regularization for recurrent neural networks was firstly introduced by Gardner in \cite{gardner_training_1989}. 
The general framework requires generating a set of $pM$ data-points from the patterns according to the following rule
\begin{equation}
    \label{eq:chi}
    S_i^{\mu,d} = \chi_i^{\mu,d}\xi_i^{\mu},\hspace{0.3cm}d = 1,..,M;\hspace{0.2cm}\mu = 1,..,p
\end{equation}
with
\begin{equation}
\small
\label{eq:Pchi}
    P(\chi_i^{\mu,d} = x) = \frac{(1+m_t)}{2}\delta(x - 1) + \frac{(1-m_t)}{2}\delta(x + 1),
\end{equation}
where $m_t$ is a \textit{training overlap}, a control parameter for the problem that sets the distance between configurations $\vec{S}^{\mu,d}$ and the relative patterns $\vec{{\xi^{\mu}}}$. Note that, according to Gardner's prescription, all variables $\chi^{\mu,d}$ are $i.i.d.$ random variables.
The final coupling matrix $\boldsymbol{J}$ results by iteration of the following learning rule:
\begin{equation}
    \label{eq:lwn2}
    J_{ij}^{(d+1)} = J_{ij}^{(d)}+ \frac{\lambda}{N}\left(\epsilon_i^{\mu,d} \xi_i^{\mu} S_j^{\mu,d} + \epsilon_j^{\mu,d} \xi_j^{\mu} S_i^{\mu,d}\right),
\end{equation}
with
\begin{equation}
    \label{eq:eps}
    \epsilon_i^{\mu,d} = \frac{1}{2}\Big(1 - \text{sign}\Big(\xi_i^{\mu}\sum_{k=1}^N J_{ik}^{(d)}S_k^{\mu,d}\Big)\Big),
\end{equation}
where we pick a pattern $\mu$ at random at each iteration $d<M$.\\
When we send $m_t \rightarrow 0^+$ this rule transforms into
\begin{equation}
    J_{ij}^{(d+1)} = J_{ij}^{(d)}-\frac{\lambda}{N}S_i^{\mu,d}S_j^{\mu,d},   
\end{equation}
which coincides with the unlearning algorithm. It was showed by \cite{wong_optimally_1990} that $\boldsymbol{J}$ converges to $\boldsymbol{J}\propto\boldsymbol{W}$ Hebbian matrix, if $\chi^{\mu,d}$ are $i.i.d.$ generated as in Eq.~\ref{eq:Pchi}. However, we proved in \cite{bene2} that $\boldsymbol{J}$ closely approaches $\boldsymbol{J}_{\text{SVM}}$ if we inject specific correlations among the entries of the $\vec{\chi}^{\mu,d}$ vector. Specifically, the correlations contained in stable fixed points of a network initialized as $\boldsymbol{J}^{(0)} = \boldsymbol{W}$ match the requirements to reproduce SVM learning. Such result explains the strong similarities between the unlearning prescription and SVM learning analyzed in \cite{benedetti_supervised_2022}.  
\subsection*{Spatial Memories: visualizing the attractors}
\begin{figure}[htbp]
\centering
\includegraphics[width=\linewidth]{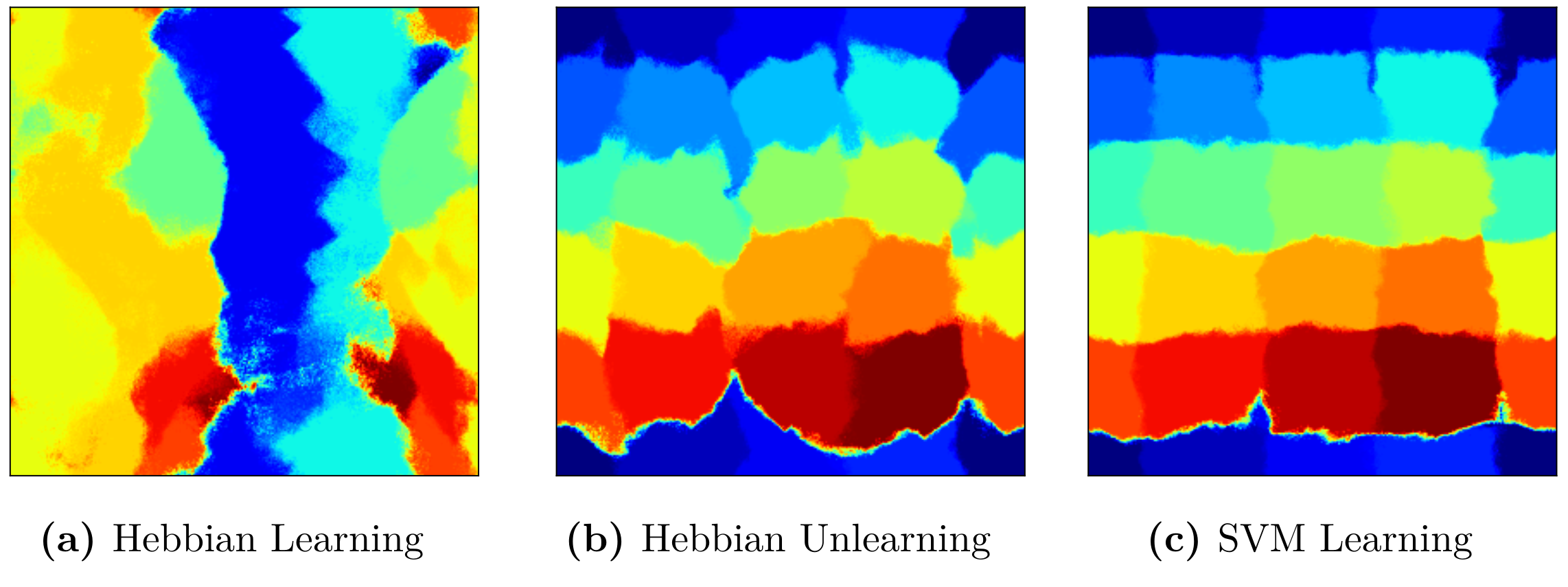}
\caption{Basins of attraction of a neural network that learns positions on a 2-dimensional grid through: Hebbian Learning, $\boldsymbol{J}=\boldsymbol{W}$  (a), Hebbian Learning \& Hebbian Unlearning (b), SVM (c). The parameters are chosen to be $N = 500$, $p = 20$, $w = 0.3$, $\lambda = 10^{-2}$. Image from \cite{prep}.}
\label{fig:space}
\end{figure}

We will now provide for a visualization of the memorization effects of the unlearning algorithm by evaluating memories with spatial correlations. We will employ the method used in \cite{battista_capacity-resolution_2020}, based on the functioning of the hyppocampus. 
Let us consider a set of $p$ positional vectors $\vec{r}^{\mu}$ living in a 2-dimensional Euclidean space $[-L,L]^2$ with periodic boundary conditions and neural configurations $\vec{\xi}^{\mu}$ from a $N$-dimensional hypercube $\{0,1\}^N$. 
In order to map the positions $\vec{r}^{\mu}$ into the neural configurations $\vec{\xi}^{\mu}$ we consider each neuron of the network to be a place-cell monitoring the space contained into a place-field of volume $f$, positioned at random in the map. Each entry $\xi_i^{\mu}$ is then set to $1$ if $\vec{r}^{\mu}$ falls under the $i$-th place field, $0$ otherwise.\\
Consider now a recurrent neural network with $N = 500$ neurons and $p=20$ spatial memories set on a 2-dimensional grid. We assign a colour to each memory. We learn the positions according to the Hebbian way, SVM and Hebbian Learning + Unlearning (taking into account the bias $f$ in the coding level, see \cite{prep} for details). Then the neural dynamics in Eq.~\ref{eq:dynamics} is initialized on many different positions in the map and run until convergence: depending on the basin of convergence, the relative colour will be assigned to the initial position. 
Figure~\ref{fig:space} depicts the shape of the basins of attraction obtained from the different learning rules. While the basins are scattered and not connected when only the Hebbian learning is performed, they tassellate the Euclidean space, and then the hypercube, when SVM and Unlearning are employed. 
This result tends to the classification scenario we have analyzed in Section~\ref{sec:class}. We conclude that Hebbian learning \& Hebbian Unlearning mimics the classification algorithm and reaches, only by seeing the memories once, quasi-maximal basins of attraction. The performance of the SVM also recalls the classification regime, in fact it is observed that moment-matching holds in SVMs (that are rather symmetric networks) when $\alpha$ is small. 

\section{\label{sec:gene}Generative Modeling}
A model is said to be \textit{generative} when it learns the probability distribution of a set of data-points, instead of the data-points themselves \cite{cocco_statistical_2022}. 
Let us call
the empirical probability distribution of the data-points $P_{data}(\vec{S})$ and the probability distribution of the model $P_{mod,J}(\vec{S})$, with $\boldsymbol{J}$ being a set of learnable parameters. One possible approach to train the model is finding the parameters $\boldsymbol{J}^*$ that minimize the Kullback-Leibler distance $D_{\text{KL}}\left(P_{mod,J}||P_{data}\right)$. 
Once the model has has fitted the statistics of the training data, we can sample, from the same distribution, brand new data-points that are indistinguishable from the training ones.
\subsection*{Boltzmann Machines}
Let us consider a set of binary data-points $\{\vec{S}^{a}\}_{a=1}^M$ with $\vec{S}^a \in \{-1,+1\}^N$. Each point can be conceived as a recurrent neural network configuration, that satisfies the Gibbs-Boltzmann statistics, i.e.
\begin{equation}
    P_{mod,J}(\vec{S}) = \frac{1}{Z}\exp{-E[\vec{S}|\boldsymbol{J}]},     
\end{equation}
with
\begin{equation}
    Z = \sum_{\vec{S}}e^{- E[\vec{S}|\boldsymbol{J}]}.
\end{equation}
Regardless the initialization of the parameters, one can perform a gradient-descent over the distance $D_{\text{KL}}\left(P_{mod,J}||P_{data}\right)$ obtaining the following updating rule for the couplings
\begin{equation}
\label{eq:BML}
    \boldsymbol{J}^{(d+1)} = \boldsymbol{J}^{(d)} + \frac{\lambda}{N}\left(\boldsymbol{W}-\boldsymbol{C}_1^{(d)}\right),
\end{equation}
with $\lambda$ learning rate, diagonal elements $J_{ii} = 0$  $\forall i$ and $\boldsymbol{W}$ being the Hebbian matrix over the training data-set. The matrix $\boldsymbol{C}_1$ collects the thermal correlations
\begin{equation}
    \boldsymbol{C}_1^{(d)} = \frac{1}{\Omega}\sum_{\nu = 1}^{\Omega} \vec{S}^{\nu}(\vec{S}^{\nu})^{\intercal},
\end{equation}
where $\vec{S}^{\nu}$ are $\Omega$ network configurations sampled at the equilibrium from the stochastic dynamics in Eq.~\ref{eq:stoch_dynamics}. The resulting model from Eq.~\ref{eq:BML} is called \textit{Boltzmann-Machine} (BM) \cite{hinton_unsupervised_1999}. 
\subsection*{Hebbian Learning and Thermal Unlearning}
As we did for adapting the algorithm in Eq.~\ref{eq:up_J} to the Unlearning procedure for associative memory (see Eq.~\ref{eq:unl_rule}), we can split the BM learning in two subsequent training phases without losing its generation performance. This means that we can perform all the Hebbian learning beforehand, by initializing $\boldsymbol{J}^{(0)}=\boldsymbol{W}$ and then the anti-Hebbian learning, by iterating: 
\begin{equation}
\label{eq:BML1}
    \boldsymbol{J}^{(d+1)} = \boldsymbol{J}^{(d)} - \frac{\lambda}{N}\boldsymbol{C}_1^{(d)}.
\end{equation}
This form of learning looks like a thermal version of the Unlearning rule evaluated in \cite{nokura_paramagnetic_1996} for memorization purposes. 
\begin{figure}[htbp]
\centering
\includegraphics[width=\linewidth]{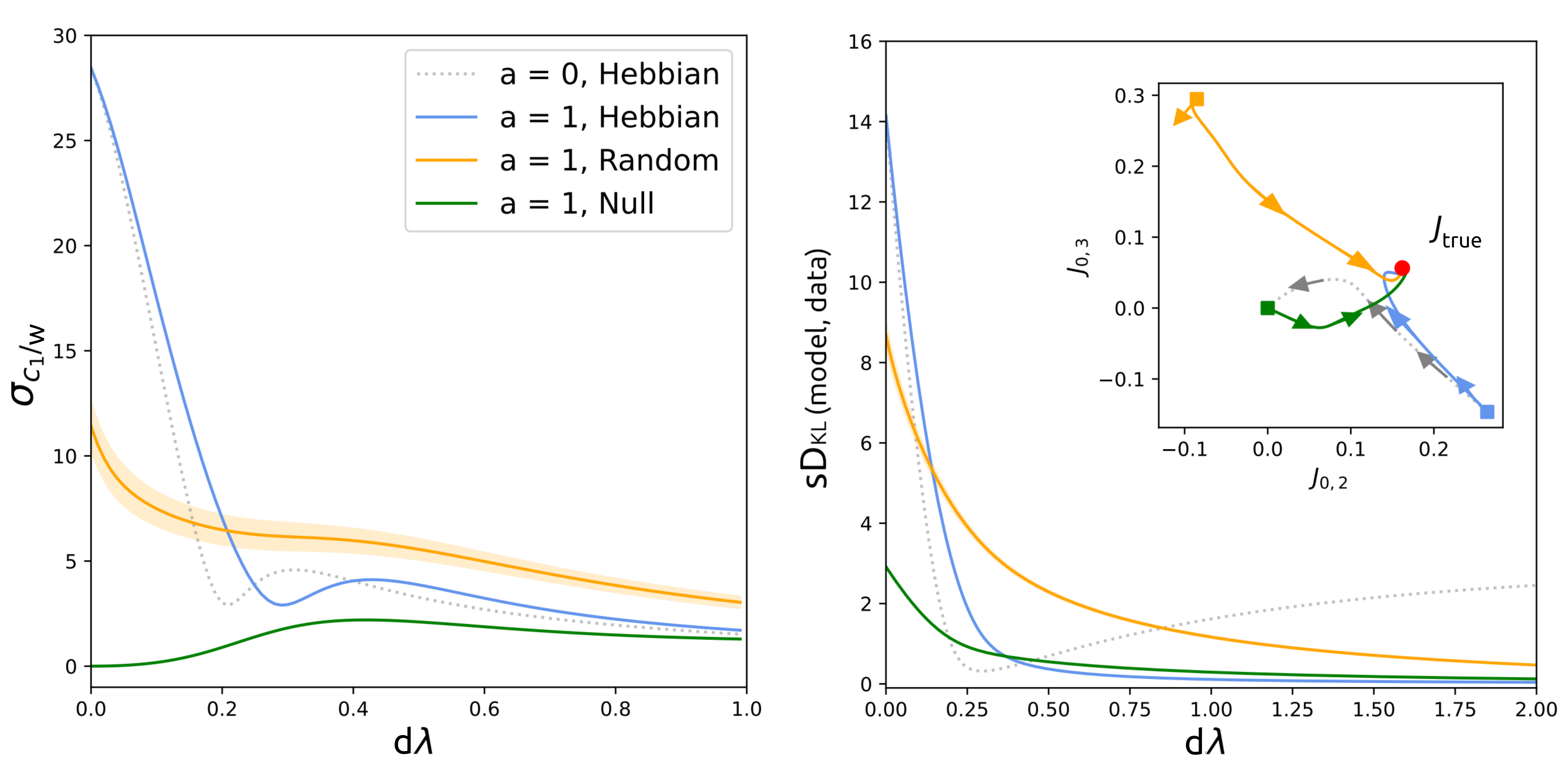}
\caption{Left: Standard deviation of the ratio between the elements of $\boldsymbol{C_1}$ and $\boldsymbol{W}$ as a function of the learning time. Right: Symmetrized Kullback-Leibler divergence between the data and the model as a function of the learning steps; the sub-panel reports a projection of the evolution of $\boldsymbol{J}$ over two picked synaptic connections: $\boldsymbol{J}_{\text{true}}$ represents the true target model. The experiment has been run on a network of size $N = 18$ so that perfect enumeration of examples was possible. Data were sampled from a SK model with the same size. Different colours of the curves are associated to different choices of the hyperparameter ($a = 0$ is unlearning, $a = 1$ is BM learning) as well as different initializations of the coupling matrix. Image adapted from \cite{ventura_2024}.}
\label{fig:th_unl}
\end{figure}
It has been numerically measured that the Hebbian initialization implieds the thermal correlations $(C_1)_{ij}$ of the system to be much larger than the data correlations $W_{ij}$. As a consequence, at least for a given amount of learning steps, the Hebbian synaptic reinforcement present in BM learning (see Eq.~\ref{eq:Juprob}) can be avoided, because Unlearning is more important. 
The left panel in Figure~\ref{fig:th_unl} reports the quantity
\begin{equation}
    \sigma_{c_1/w} = \sqrt{\frac{2}{N(N-1)}\sum_{i,j>i}\left(\frac{(C_1)_{ij}}{W_{ij}}\right)^2},
\end{equation}
as a function of the learning time for BM with different types of initializations of the pararameters (Hebbian, Random and All-zero connections) and Thermal Unlearning. The experiments were run in small system learning configurations generated from a spin-glass. It results that the Hebbian initialization favours high values of $\sigma_{c_1/w}$ up to a certain amount of training steps, after which this quantity becomes $\mathcal{O}(1)$. As reported in the right panel of Figure~\ref{fig:th_unl}, this effect implies a good agreement between BM learning and the new procedure in descending the KL distance between the model and the data. Yet again, this holds up to an amount of training steps, after which the unlearning performance decays: the separation of the Hebbian and anti-Hebbian learning phases always carries the necessity of introducing and early-stopping criterion, which is discussed in \cite{benedetti_supervised_2022,ventura_2024,prep}.  
\subsection*{Unlearning avoids Criticality}
Past literature \cite{tkacik_thermodynamics_2015,mora_are_2011} highlighted that critical models, i.e. inferred systems being highly susceptible to small changes in the parameters, can be attractive for BM learning. In fact, if we consider training as an homogeneous sampling of the models that minimize $D_{\text{KL}}\left(P_{mod,J}||P_{data}\right)$, critical models have a large basin of attraction and are thus sampled most often~\cite{mastromatteo_criticality_2011}.
Criticality can be problematic for data generation: it implies a long correlation time in the Monte Carlo dynamics, slowing down the sampling process; it makes the model susceptible under rescaling of the parameters, reducing its predictivity in real data applications. Hence, avoiding criticality might help increasing generalization.
\\We propose a regularization to the model \cite{ventura_2024} that consists in minimizing the following quantity
\begin{multline}
    \label{eq:obj}
      \mathcal{L}_a(\boldsymbol{J}) = D_{\text{KL}}\left(P_{mod,J}||P_{data}\right) \\ + \left(\frac{a-1}{a}\right) D_{\text{KL}}\left(P_{mod,aJ}||P_{data}\right),
\end{multline}
which translates into having the following gradient-descent equations 
\begin{equation}
    \label{eq:Juprob}
    \small
    \boldsymbol{J}^{(d+1)} = \boldsymbol{J}^{(d)} +\frac{\lambda}{N}\big[ a\left(\boldsymbol{W} - \boldsymbol{C}_a^{(d)}\right) - (\boldsymbol{C}_1^{(d)} - \boldsymbol{C}_a^{(d)})\big].
\end{equation}
Notice that Eq.~\ref{eq:Juprob} coincides with Eq.~\ref{eq:BML} when $a = 1$ and becomes equal to Eq.~\ref{eq:BML1} when $a = 0$. 
The scope of the regularization in Eq.~\ref{eq:obj} is to shift the model at $T = 1$, from which new examples are sampled, towards a state of the system where correlations are lower. This effect can be measured by computing the specific heat $C_v(T)$ of the system, that represents the susceptibility of the model under a small rescaling of the parameters $\boldsymbol{J}$.
\begin{figure}[htbp]
\centering
\includegraphics[width=0.9\linewidth]{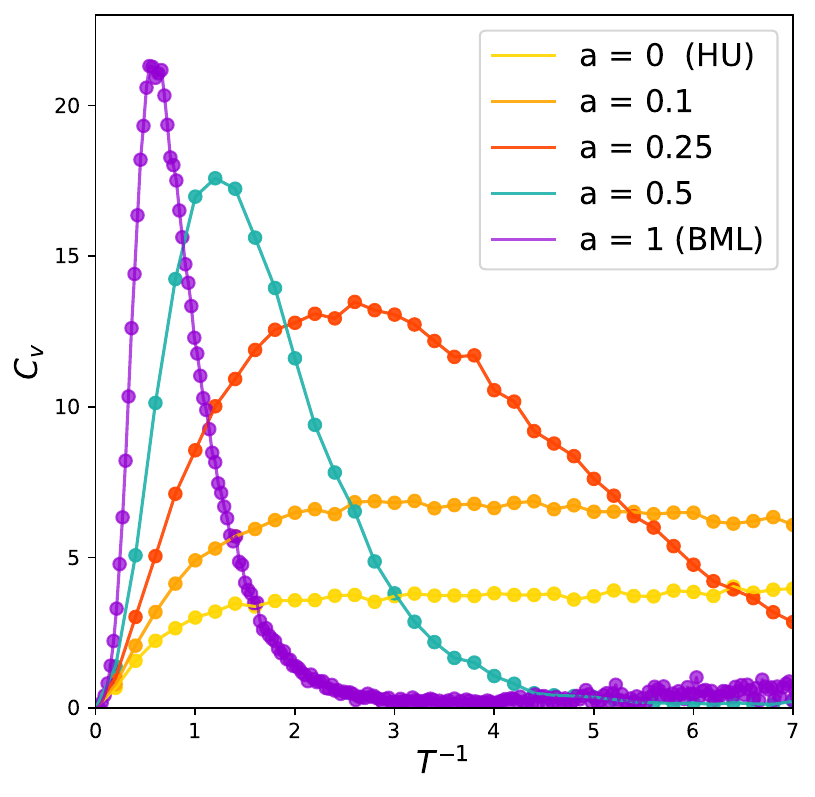}
\caption{Specific heat $C_v$ as a function of the inverse temperature $1/T$ for the model inferred from neuronal data collected in \cite{schneidman_weak_2006}. The unlearning regularization is applied with different choices of the hyperparameter $a$. Image adapted from \cite{ventura_2024}.}
\label{fig:bialek}
\end{figure}
The effect of the Unlearning regularization on the robustness of the model under rescaling of the parameters is showed in Figure~\ref{fig:bialek}. In the plot, real neuronal data from \cite{schneidman_weak_2006} have been learned by the model with different choices of $a$ wich $a \leq 1$. 
An early stopping criterion, described in \cite{ventura_2024} has been employed to guarantee the best agreement between the data and the model. As noticeable from the experiment, the curve of the specific heat flattens when $a \rightarrow 0$, suggesting an improvement in the robustness of the model and the avoidance of criticality. 
\section{Conclusions and Open Questions}
The main goal of our analysis is to show that a general learning rule for neural networks can be adapted to classify data-points, memorize them, and learn their probability distribution to generate new examples. In the case of memorization and generation, this learning prescription can be split in two phases: a Hebbian reinforcement phase followed by an anti-Hebbian regularization phase, also called Unlearning. 
This rule is said to be \textit{local}, because it updates the connections of the network based on the activity of the single pairs of neurons and not on the entire system. 
This rule is also \textit{unsupervised}, because it relies on the covariance of the data-point and not on the data-points themselves; moreover, once we consider the Hebbian phase as separated from the Unlearning one, the model sees the data only once, at initialization, while the rest of the process does not consult them at all. 
Let us conclude by resuming what are the main contributions and open questions led by the Learning \& Unlearning algorithm in the fields of Biology, Computer Science and Physics.\\
\subsection*{Biology}
The Learning and Unlearning principle has emerged in different forms in the past neuroscience and biophysics literature and this paper might finally try to connect them. It was 1983: while Crick and Mitchison \cite{crick_function_1983} conjectured a \textit{pruning} role REM sleep in the synaptic network, Hopfield was proposing a formal algorithm that reproduced this mechanism through its Unlearning routine \cite{hopfield_unlearning_1983}. 
A few years later Földiák and Barlow \cite{foldiak_forming_1990, barlow_adaptation_1989} wrote about the sparse regularization role of anti-Hebbian learning in associative memory models, while never mentioning previous Hopfield's Unlearning procedure. At the end of the '90s, theoretical neuroscientists proposed different models of homeostatic rules \cite{turrigiano_homeostatic_1999}, following experimental observations in synaptic plasticity. The contemporary work from \cite{horn} discusses the formal similarities between Unlearning and Homeostatic regulation. Eventually, the more recent Synaptic Homeostasis Hypothesis  (SHY) \cite{tononi,girardeau_brain_2021} in neuroscience supports initial Crick and Mitchison's ideas that learning might be composed of a Hebbian learning phase during wakefulness, and a subsequent offline phase of pruning and regularization of the synapses taking place during sleep. This last phase would have the homeostatic role of conserving the total daily synaptic volume.
Even if obtaining a direct measure of synaptic plasticity in real systems is still not very much affordable, our analysis corroborates the possibility of finding, at least indirectly (e.g. through inferential studies), anti-Hebbian mechanisms in the brain responsible for memory reinforcement and generative capabilities.   
\subsection*{Computer Science}
As recently highlighted by Hinton by \cite{hinton_forward-forward_2022}, a moment-matching rule as the one under analysis might help even feed-forward generalizations of recurrent neural networks to avoid back-propagation methods and being trained faster in an unsupervised fashion. \\
A few works in literature \cite{pozas-kerstjens_efficient_2021}, apart from ours \cite{ventura_2024}, investigate the importance of Hebbian initialization as a good start for training neural networks. We do believe that deepening this point, along properties related to the data-structure, would be an important research topic for the future.  
\subsection*{Physics}
Exploring these types of models may sometimes require a deeper investigation into the statistical mechanics itself. One question arises from the classification problem in Section~\ref{sec:class}. The limitations of the algorithm occurring at higher values of $N$ might be overcome by the choice of a different dynamics. A big open question in Stat-Mech is finding dynamic rules that are able to probe the deepest states in glassy landscapes \cite{erba,riccardo}, and our problem might be an interesting case study.
It is worth to notice that the classification problem evaluated in Section~\ref{sec:class} can be mapped into a extensive-rank matrix factorization problem, which is the central object of future publications \cite{prep}.
Another interesting topic that deserves more effort from the community definitely is the glassy phase of Hebbian networks (i.e. the Hopfield model), as a function of $\alpha > 0.14$. It is clear that glassy states and saddles in the landscape inherit interesting topological properties by the memories, even if they are no more stable themselves, implying the Hebbian network to be a good initialization for neural systems.

\section*{Acknowledgments}
The author thanks the organizers of the IEEE conference COMPENG2024 as well as the many partecipants for exchanging ideas and discussions. The author is also devoted to his mentors and collaborators from the previous three years of research, in particular: Marco Benedetti, Simona Cocco, Carlo Lucibello, Enzo Marinari, Rémi Monasson, Giancarlo Ruocco and Francesco Zamponi. Fundings were provided by Next Generation EU PRIN 2022 CUP J53D23001330001.  

\bibliographystyle{unsrt}
\bibliography{biblio}

\begin{thebibliography}{10}

\bibitem{amit_modeling_nodate}
D.J. Amit.
\newblock {\em Modeling {Brain} {Function}}.
\newblock Cambridge University Press, 1989.

\bibitem{hebb_organization_1949}
D.O. Hebb.
\newblock {\em The {Organization} of {Behavior} : {A} {Neuropsychological} {Theory}}.
\newblock John Wiley and Sons, 1949.

\bibitem{hopfield_neural_1982}
J.J. Hopfield.
\newblock Neural networks and physical systems with emergent collective computational abilities.
\newblock {\em Proceedings of the National Academy of Sciences}, 79(8):2554, 1982.

\bibitem{gardner_space_1988}
E.~Gardner.
\newblock The space of interactions in neural network models.
\newblock {\em Journal of Physics A: Mathematical and General}, 21(1):257, 1988.

\bibitem{hinton_unsupervised_1999}
J.~Hinton and T.J. Sejnowski.
\newblock {\em Unsupervised {Learning}: {Foundations} of {Neural} {Computation}}.
\newblock The MIT Press, 1999.

\bibitem{prep}
E.~Ventura.
\newblock Factorizing extensive-rank matrices while building optimal associative memories.
\newblock {\em in Preparation}, 2024.

\bibitem{hopfield_unlearning_1983}
J.J. Hopfield, D.I. Feinstein, and R.G. Palmer.
\newblock ‘{Unlearning}’ has a stabilizing effect in collective memories.
\newblock {\em Nature}, 304(5922):158, 1983.

\bibitem{gardner_phase_1989}
E.~Gardner, H.~Gutfreund, and I.~Yekutieli.
\newblock The phase space of interactions in neural networks with definite symmetry.
\newblock {\em Journal of Physics A: Mathematical and General}, 22(12):1995, 1989.

\bibitem{gardner_optimal_1989}
E.~Gardner.
\newblock Optimal basins of attraction in randomly sparse neural network models.
\newblock {\em Journal of Physics A: Mathematical and General}, 22(12):1969, 1989.

\bibitem{benedetti_supervised_2022}
M.~Benedetti, E.~Ventura, E.~Marinari, G.~Ruocco, and F.~Zamponi.
\newblock Supervised perceptron learning vs unsupervised {Hebbian} unlearning: {Approaching} optimal memory retrieval in {Hopfield}-like networks.
\newblock {\em The Journal of Chemical Physics}, 156(10):104107, 2022.

\bibitem{bene2}
M.~Benedetti and E.~Ventura.
\newblock Training neural networks with structured noise improves classification and generalization.
\newblock {\em arXiv:2302.13417}, 2024.

\bibitem{ventura_2024}
E.~Ventura, S.~Cocco, R.~Monasson, and F.~Zamponi.
\newblock Unlearning regularization for {Boltzmann} {Machines}.
\newblock {\em Machine Learning: Science and Technology}, 5:025078, 2024.

\bibitem{peretto_collective_1984}
P.~Peretto.
\newblock Collective properties of neural networks: a statistical physics approach.
\newblock {\em Biological Cybernetics}, 50(1):51--62, 1984.

\bibitem{mezard_spin_1986}
M.~Mezard, G.~Parisi, and M.~Virasoro.
\newblock {\em Spin {Glass} {Theory} and {Beyond}: {An} {Introduction} to the {Replica} {Method} and {Its} {Applications}}, volume~9 of {\em World {Scientific} {Lecture} {Notes} in {Physics}}.
\newblock World Scientific, 1986.

\bibitem{bengio}
Y.~Bengio, A.~Courville, and P.~Vincent.
\newblock Representation learning: A review and new perspectives.
\newblock {\em IEEE Transactions on Pattern Analysis and Machine Intelligence}, 35(8):1798--1828, 2013.

\bibitem{amit_storing_1985}
D.~J. Amit, H.~Gutfreund, and H.~Sompolinsky.
\newblock Storing {Infinite} {Numbers} of {Patterns} in a {Spin}-{Glass} {Model} of {Neural} {Networks}.
\newblock {\em Physical Review Letters}, 55(14):1530, 1985.

\bibitem{brunel_is_2016}
N.~Brunel.
\newblock Is cortical connectivity optimized for storing information?
\newblock {\em Nature Neuroscience}, 19(5):749--755, 2016.

\bibitem{battista_capacity-resolution_2020}
A.~Battista and R.~Monasson.
\newblock Capacity-resolution trade-off in the optimal learning of multiple low-dimensional manifolds by attractor neural networks.
\newblock {\em Physical Review Letters}, 124(4):048302, 2020.

\bibitem{van_hemmen_increasing_1990}
J.L. Van~Hemmen, L.B. Ioffe, R.~Kühn, and M.~Vaas.
\newblock Increasing the efficiency of a neural network through unlearning.
\newblock {\em Physica A: Statistical Mechanics and its Applications}, 163(1):386, 1990.

\bibitem{gardner_training_1989}
E.~J. Gardner, D.~J. Wallace, and N.~Stroud.
\newblock Training with noise and the storage of correlated patterns in a neural network model.
\newblock {\em Journal of Physics A: Mathematical and General}, 22(12):2019, 1989.

\bibitem{wong_optimally_1990}
K.Y.M. Wong and D.~Sherrington.
\newblock Optimally adapted attractor neural networks in the presence of noise.
\newblock {\em Journal of Physics A: Mathematical and General}, 23(20):4659, 1990.

\bibitem{cocco_statistical_2022}
S.~Cocco, R.~Monasson, and F.~Zamponi.
\newblock {\em From {Statistical} {Physics} to {Data}-driven {Modelling}: {With} {Applications} to {Quantitative} {Biology}}.
\newblock Oxford University Press, 2022.

\bibitem{nokura_paramagnetic_1996}
K.~Nokura.
\newblock Paramagnetic unlearning in neural network models.
\newblock {\em Physical Review E}, 54(5):5571, 1996.

\bibitem{tkacik_thermodynamics_2015}
G.~Tkačik, T.~Mora, O.~Marre, D.~Amodei, S.~E. Palmer, M.~J. Berry, and W.~Bialek.
\newblock Thermodynamics and signatures of criticality in a network of neurons.
\newblock {\em Proceedings of the National Academy of Sciences}, 112(37):11508--11513, 2015.

\bibitem{mora_are_2011}
T.~Mora and W.~Bialek.
\newblock Are {Biological} {Systems} {Poised} at {Criticality}?
\newblock {\em Journal of Statistical Physics}, 144(2):268--302, 2011.

\bibitem{mastromatteo_criticality_2011}
I.~Mastromatteo and M.~Marsili.
\newblock On the criticality of inferred models.
\newblock {\em Journal of Statistical Mechanics: Theory and Experiment}, 2011(10):P10012, 2011.

\bibitem{schneidman_weak_2006}
E.~Schneidman, M.~Berry, R.~Segev, and W.~Bialek.
\newblock Weak pairwise correlations imply strongly correlated network states in a neural population.
\newblock {\em Nature}, 440(7087):1007--1012, 2006.

\bibitem{crick_function_1983}
F.~Crick and G.~Mitchison.
\newblock The function of dream sleep.
\newblock {\em Nature}, 304(5922):111, 1983.

\bibitem{foldiak_forming_1990}
P.~Földiák.
\newblock Forming sparse representations by local anti-{Hebbian} learning.
\newblock {\em Biological Cybernetics}, 64(2):165--170, 1990.

\bibitem{barlow_adaptation_1989}
H.~Barlow and P.~Földiák.
\newblock Adaptation and decorrelation in the cortex.
\newblock In {\em The Computing Neuron}. Addison-Wesley Longman Publishing Co., Inc., 1989.

\bibitem{turrigiano_homeostatic_1999}
G.~Turrigiano.
\newblock Homeostatic plasticity in neuronal networks: the more things change, the more they stay the same.
\newblock {\em Trends in Neurosciences}, 22(5):221--227, 1999.

\bibitem{horn}
D.~Horn, N.~Levy, and E.~Ruppin.
\newblock Neuronal regulation vs synaptic unlearning in memory maintenance mechanisms.
\newblock {\em Network}, 9(4):577--586, 1998.

\bibitem{tononi}
G.~Tononi and C.~Cirelli.
\newblock Sleep function and synaptic homeostasis.
\newblock {\em Sleep Medicine Reviews}, 10:49--62, 2006.

\bibitem{girardeau_brain_2021}
G.~Girardeau and V.~Lopes-dos Santos.
\newblock Brain neural patterns and the memory function of sleep.
\newblock {\em Science}, 374(6567):560--564, 2021.

\bibitem{hinton_forward-forward_2022}
G.~Hinton.
\newblock The {Forward}-{Forward} {Algorithm}: {Some} {Preliminary} {Investigations}.
\newblock {\em arXiv:2212.13345}, 2022.

\bibitem{pozas-kerstjens_efficient_2021}
A.~Pozas-Kerstjens, G.~Muñoz-Gil, E.~Piñol, M.~A. García-March, A.~Acín, M.~Lewenstein, and P.~R. Grzybowski.
\newblock Efficient training of energy-based models via spin-glass control.
\newblock {\em Machine Learning: Science and Technology}, 2(2):025026, 2021.

\bibitem{erba}
E.~Vittorio, F.~Behrens, F.~Krzakala, and L.~Zdeborová.
\newblock Quenches in the sherrington-kirkpatrick model.
\newblock {\em arXiv:2405.04267v1}, 2024.

\bibitem{riccardo}
R.~Zecchina.
\newblock Neural network with local short memory and complex updating.
\newblock {\em International Journal of Neural Systems}, 3(4):379--387, 1992.

\end{thebibliography}


\end{document}